# Tunable Γ − K Valley Populations in Hole-Doped Trilayer WSe₂

Hema C. P. Movva,[1] Timothy Lovorn,[2] Babak Fallahazad,[1] Stefano Larentis,[1] Kyounghwan Kim,[1]
Takashi Taniguchi,[3] Kenji Watanabe,[3] Sanjay K. Banerjee,[1] Allan H. MacDonald,[2] and Emanuel Tutuc[1, *]

[1]*Microelectronics Research Center, Department of Electrical and Computer Engineering,*
*The University of Texas at Austin, Austin, TX 78758, USA*
[2]*Department of Physics, The University of Texas at Austin, Austin, TX 78712, USA*
[3]*National Institute of Materials Science, 1-1 Namiki Tsukuba, Ibaraki 305-0044, Japan*
(Dated: January 7, 2018)

We present a combined experimental and theoretical study of valley populations in the valence bands of trilayer WSe₂. Shubnikov−de Haas oscillations show that trilayer holes populate two distinct subbands associated with the $K$ and $\Gamma$ valleys, with effective masses $0.5m_e$ and $1.2m_e$, respectively; $m_e$ is the bare electron mass. At a fixed total hole density, an applied transverse electric field transfers holes from $\Gamma$ orbitals to $K$ orbitals. We are able to explain this behavior in terms of the larger layer polarizability of the $K$ orbital subband.

Transition metal dichalcogenides (TMDs) are layered materials that possess strong spin-orbit coupling, have large carrier effective masses, and can be isolated down to a monolayer, making them attractive hosts for phenomena associated with strong spin-orbit and electron-electron interactions. Symmetry considerations dictate that monolayer 2H−TMDs possess band extrema at the time-reversed partner $K$ and $K'$ Brillouin zone corners with a finite gap between dipole coupled conduction and valence bands [1]. Beyond the monolayer limit, the TMD bandstructure becomes more complicated. In particular, the locations of the band extrema in few-layer TMDs [2, 3], and their dependence on externally controllable parameters like transverse gate electric fields ($E$), have, for most TMDs [4–7], remained open questions that can be answered only by combining experiment and theory.

Here we address the band maxima and impact of an $E$-field in the valence band of trilayer WSe₂, a TMD with large valence band spin-orbit splitting [8], high-mobility, and robust low temperature Ohmic contacts [9–11]. Recent studies have shown that the valence band maxima are located at the $K$ points in mono and bilayer WSe₂, and at the $\Gamma$ point in bulk WSe₂ [3, 12]. Using magnetotransport measurements, we show that in high mobility WSe₂ trilayers encapsulated in $h$-BN, holes populate two distinct subbands with different effective masses, $0.5m_e$ and $1.2m_e$, that we associate with the $K$ and $\Gamma$ valleys respectively; $m_e$ is the bare electron mass. At a fixed total hole density, the $K$ and $\Gamma$ occupations can be tuned by an applied $E$-field in a dual-gated device, with $\Gamma$ being the lowest energy state at low $E$-field and $K$ being the lowest energy state at high $E$-field. *Ab-initio* calculations support these findings, and explain the shift of the valence band maxima, and the consequent transfer of holes from $\Gamma$ to $K$ with increasing $E$.

Our samples consist of WSe₂ trilayers exfoliated from bulk crystals (HQ Graphene). Figure 1(a) shows the room temperature photoluminescence (PL) spectrum of trilayer 2H−WSe₂ which has distinct peaks corresponding to the indirect (1.45 eV) and direct (1.60 eV) energy gap transitions [13]. The PL spectrum combined with optical contrast and Raman spectroscopy enables an unambiguous identification of trilayer WSe₂. Figure 1(a) inset shows the optical micrograph of an

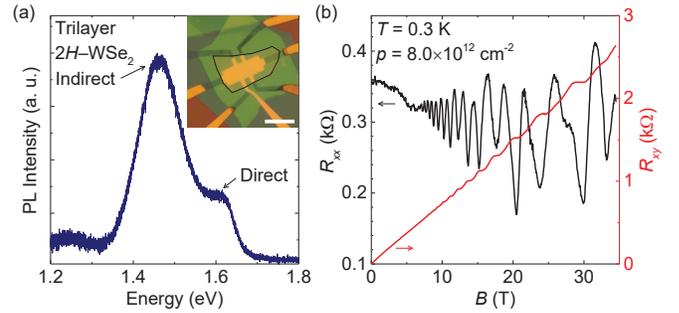

FIG. 1. (a) Room temperature PL spectrum of trilayer 2H−WSe₂ acquired with 532 nm wavelength excitation. The peaks corresponding to the direct and indirect gap transitions are labeled. Inset: Optical micrograph of a dual-gated trilayer WSe₂ Hall bar. The black line marks the contour of the WSe₂ flake. The scale bar is 10 $\mu$m. (b) Trilayer WSe₂ $R_{xx}$ and $R_{xy}$ vs $B$ measured at $T = 0.3$ K and $p = 8.0 \times 10^{12}$ cm$^{-2}$.

$h$-BN encapsulated, dual-gated trilayer WSe₂ Hall bar sample, fabricated using a van der Waals assembly technique [9, 14]. Ohmic hole contacts down to cryogenic temperatures were achieved using bottom Pt electrodes in combination with a large, negative top-gate bias ($V_{TG}$) [9, 10]. The dual-gated device structure allows independent control of the WSe₂ hole carrier density ($p$) and the $E$-field. Magnetotransport measurements were conducted using low frequency lock-in techniques at temperatures down to $T = 0.3$ K, and perpendicular magnetic fields up to $B = 35$ T. Three samples were investigated in this study, all with consistent results. Here we focus on data from two samples. Figure 1(b) shows the longitudinal ($R_{xx}$) and Hall ($R_{xy}$) resistance vs $B$ for a trilayer WSe₂ sample at $V_{TG} = -8.5$ V, and a back-gate bias, $V_{BG} = 0$ V. The hole density is $p = 8.0 \times 10^{12}$ cm$^{-2}$, as extracted from the slope of $R_{xy}$ at low $B$-fields. The $R_{xx}$ data show well-defined Shubnikov−de Haas (SdH) oscillations accompanied by the emergence of quantum Hall state plateaus in $R_{xy}$ coincident with $R_{xx}$ minima at high $B$-fields. Unlike holes in mono and bilayer WSe₂ at $V_{BG} = 0$ V which show one fundamental SdH oscillation frequency, corresponding to a single populated subband [10], the trilayer data of Fig. 1(b) show a clear beating

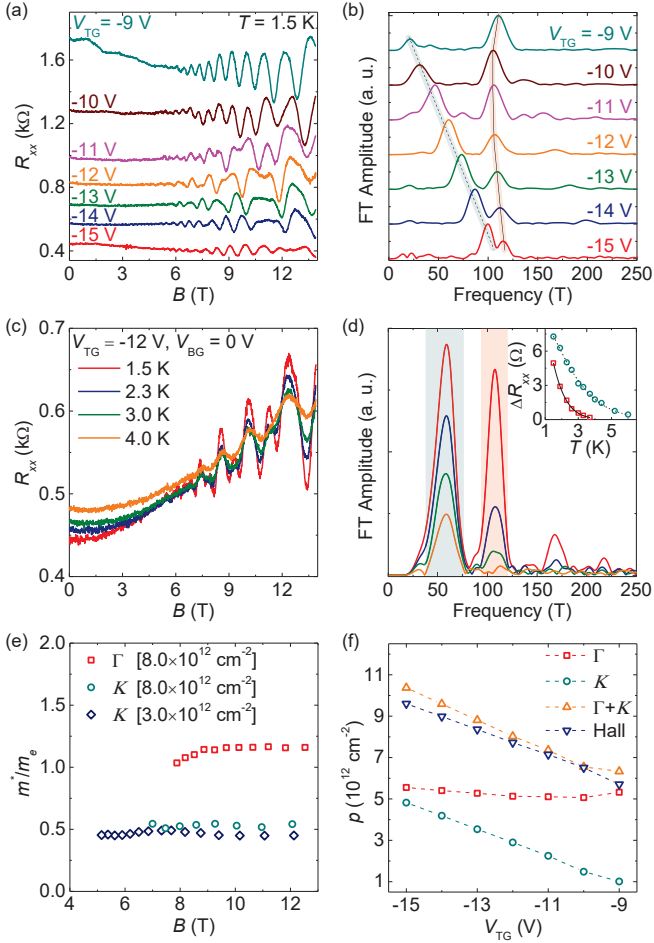

FIG. 2. (a) $R_{xx}$ vs $B$ at various $V_{TG}$ values, $V_{BG} = 0$ V, and $T = 1.5$ K. (b) FT amplitude of the $R_{xx}$ vs $B^{-1}$ data of panel (a). The traces are shifted for clarity in panels (a) and (b). The solid (dashed) line tracks the high (low)-frequency peak in panel (b). (c) $R_{xx}$ vs $B$ measured at $V_{TG} = -12$ V, $V_{BG} = 0$ V, and at different $T$ values. (d) FT amplitude of the $R_{xx}$ vs $B^{-1}$ data of panel (c). Inset: $\Delta R_{xx}$ vs $T$ at $B = 8.5$ T calculated from the inverse FT of panel (d) data, using band pass filters (shaded bands). The dashed (solid) line shows the Dingle factor fit to the low (high)-frequency peak. (e) $m^*/m_e$ vs $B$ at $V_{TG} = -12$ V, $V_{BG} = 0$ V [$p = 8.0 \times 10^{12}$ cm$^{-2}$], and at $V_{TG} = -12$ V, $V_{BG} = 6$ V [$p = 3.0 \times 10^{12}$ cm$^{-2}$]. The light (heavy) holes belong to the $K$ ($\Gamma$) valley. (f) $p$ vs $V_{TG}$ associated with the individual FT peaks of panel (b) data, their sum, and Hall measurements.

pattern, suggestive of holes populating multiple subbands.

To investigate the origin of the SdH oscillations beating pattern, we performed magnetotransport measurements as a function of $V_{TG}$ and $V_{BG}$. Figure 2(a) shows $R_{xx}$ vs $B$ measured in a second trilayer WSe$_2$ sample with a top (back)-gate capacitance, $C_{TG} = 120$ nF/cm$^2$ ($C_{BG} = 130$ nF/cm$^2$), at different $V_{TG}$ values, $V_{BG} = 0$ V, and $T = 1.5$ K. Figure 2(b) shows the Fourier transform (FT) amplitude of the $R_{xx}$ vs $B^{-1}$ data of Fig. 2(a), which shows two principal peaks that differ markedly in their response to $V_{TG}$. While the low-frequency peak increases in frequency with $|V_{TG}|$, the high-frequency peak position is nearly independent of $V_{TG}$. A similar set of $R_{xx}$ vs $B$ data at different $V_{BG}$ values, and fixed $V_{TG}$ is discussed in the Supplemental Material [15]. The presence of two principal SdH frequency peaks in the FT data suggest the existence of holes at the Fermi level in two distinct subbands. The dependence of the two FT peaks on $V_{TG}$ in Fig. 2(b) suggests that the two subbands have very different spatial confinement properties.

To probe the nature of the subbands further we measured the SdH oscillations $T$-dependence. Figure 2(c) shows $R_{xx}$ vs $B$ at $V_{TG} = -12$ V, $V_{BG} = 0$ V, at various $T$ values, and Fig. 2(d) the corresponding FT data. The FT amplitudes of the two peaks show distinct $T$-dependence, with the high-frequency peak decaying more rapidly with $T$ compared to the low-frequency peak. This observation is indicative of different effective masses for holes in the two subbands. We extracted the individual effective mass ($m^*$) associated with each subband as follows. First, we isolated the FT peak of interest by applying a band pass filter centered at its peak and performed an inverse FT. We then performed a Dingle factor fit to each peak's inverse FT oscillations amplitude ($\Delta R_{xx}$) vs $T$ data at a fixed $B$-field, $\Delta R_{xx} \propto \xi / \sinh \xi$, where $\xi = 2\pi^2 k_B T / \hbar \omega_c$ and $\omega_c = eB/m^*$ [Fig. 2(d) inset]; $\hbar$ is the reduced Planck's constant, and $k_B$ is the Boltzmann constant. Figure 2(e) summarizes the extracted $m^*$ vs $B$ at two densities. At $V_{TG} = -12$ V, $V_{BG} = 0$ V, corresponding to $p = 8.0 \times 10^{12}$ cm$^{-2}$, holes associated with the low (high)-frequency subband have an $m^* = 0.5 m_e$ ($m^* = 1.2 m_e$). The lower $m^*$ matches closely with $m^* = 0.45 m_e$ measured for $K$ valley holes in mono and bilayer WSe$_2$ [10], while the larger $m^*$ is closer to $m^* = 0.89 m_e$ measured for $\Gamma$ valley holes in few-layer WSe$_2$ [16]. We therefore assign the light (heavy) holes to the $K$ ($\Gamma$) valley of trilayer WSe$_2$. At $V_{TG} = -12$ V, $V_{BG} = 6$ V [$p = 3.0 \times 10^{12}$ cm$^{-2}$], only $m^* = 0.5 m_e$ peaks are observed, implying that only the $K$ valley is populated [15].

To substantiate this interpretation, we compare the Hall densities with the densities associated with the FT peaks frequency ($f$), $p = g \times e/h \times f$, where $g$ is the Landau level (LL) degeneracy of the subband associated with $f$, $e$ is the electron charge, and $h$ is Planck's constant. Figure 2(f) shows $p$ vs $V_{TG}$ determined from Hall measurements, along with the individual $K$ and $\Gamma$ densities, and their sum ($\Gamma + K$) determined from the two FT peaks of Fig. 2(b) data using $g = 2$. The close match between the Hall and $\Gamma + K$ densities validates our valley designation. The $g = 2$ LL degeneracy for holes in both $\Gamma$ and $K$ valleys in WSe$_2$ is consistent with previous magnetotransport studies [10, 16].

To assess the impact of the transverse $E$-field on the $\Gamma$ and $K$ valley densities, we performed magnetotransport measurements as a function of $E = |C_{TG}V_{TG} - C_{BG}V_{BG}|/2\epsilon_0$, and at constant total density; $\epsilon_0$ is the vacuum permittivity. Figure 3(a) shows $R_{xx}$ vs $B$ at a series of $E$ values, for constant $p = 8.0 \times 10^{12}$ cm$^{-2}$, and $T = 1.5$ K. The $R_{xx}$ SdH oscillations beating pattern changes with $E$-field, suggesting an $E$-dependence of the relative valley occupations. Figure 3(b) shows the FT data associated with Fig. 3(a); the $\Gamma$ ($K$) peak frequency and the corresponding density decreases (increases) with increasing $E$. Fig. 3(c) summarizes the individual $\Gamma$, $K$ valley densities vs $E$ extracted from the FT peaks of Fig. 3(b). At low $E$-fields, a majority of the holes reside in the $\Gamma$ valley, and progressively get transferred to the $K$ valley with increasing $E$. This observation demonstrates that the $\Gamma - K$ splitting in the valence band of trilayer WSe$_2$ is $E$-field-dependent.

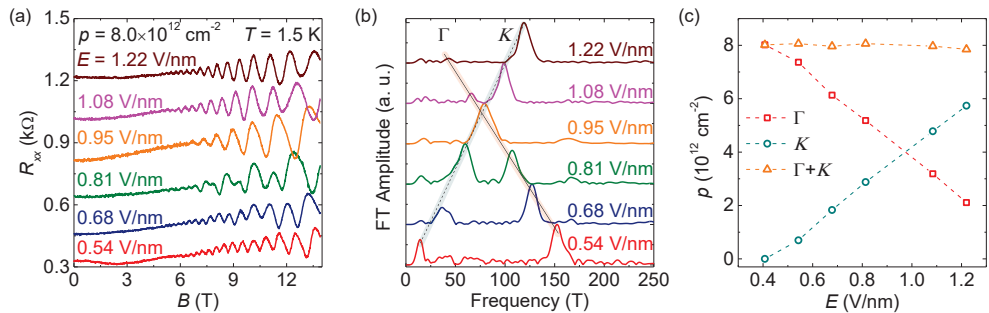

FIG. 3. (a) $R_{xx}$ vs $B$ at various $E$ values, $p = 8.0 \times 10^{12}$ cm$^{-2}$, and $T = 1.5$ K. (b) FT amplitude of the $R_{xx}$ vs $B^{-1}$ data of panel (a). The traces are shifted for clarity in panels (a) and (b). The $\Gamma$ ($K$) peak frequency decreases (increases) with increasing $E$. The solid and dashed lines are guides to the eye. (c) $p$ vs $E$ associated with the $\Gamma$ and $K$ valleys show a linear decrease (increase) of the $\Gamma$ ($K$) valley density with increasing $E$. The sum ($\Gamma + K$) is $p = 8 \times 10^{12}$ cm$^{-2}$. The two peaks overlap at $E = 0.95$ V/nm.

To understand the origin of the $E$-field-dependent $\Gamma - K$ splitting, we performed fully-relativistic density functional theory (DFT) calculations for the trilayer $2H$–WSe$_2$ system [15] under the local-density approximation [17], using the Quantum Espresso distribution [18]. As shown in Fig. 4(a), we find that the valence band maximum at the $K$ point responds much more strongly to the $E$-field than the valence band maximum at $\Gamma$. A small $E$-field is sufficient to induce substantial localization of the three high energy $K$ subbands in the individual WSe$_2$ layers, leading to a threefold splitting of these states, which increases nearly linearly with $E$, shown in Fig. 4(b). Tracking the difference between the valence band maxima $E_\Gamma - E_K$, shown by the red squares in Fig. 4(c), we find that the highest-energy valence band shifts from $\Gamma$ to $K$ for $E \gtrsim 0.9$ V/nm.

From the DFT calculation at each $E$ value, we extract a tight-binding model for the valence and conduction bands near the Fermi level using Wannier90 [19]. From each tight-binding model, we then construct $\mathbf{k} \cdot \mathbf{p}$ models for states in the $K$ and $\Gamma$ valleys, defined on a basis of two states in each WSe$_2$ layer, capturing the six highest valence bands. The $E$-dependence of the layer "on-site" energies yields an effective dielectric constant, $\epsilon_{r,v}^* = 2eEd/(E_{v,top} - E_{v,bot})$; $d$ is the displacement of each WSe$_2$ layer, i.e., the transverse distance between W atoms, and $E_{v,top}$ and $E_{v,bot}$ are the on-site energies for states in the top and bottom layers in valley $v = K$ or $\Gamma$. Averaging the values obtained at $K$ and $\Gamma$ for $E$ from 0.5 V/nm to 1.2 V/nm, we find an effective dielectric constant, $\epsilon_r^* = 7.87$, comparable to the value of 7.2 previously observed in multilayer WSe$_2$ [20]. Using our value of $\epsilon_r^*$, we can generate a $\mathbf{k} \cdot \mathbf{p}$ model for any $E$ from the $\mathbf{k} \cdot \mathbf{p}$ model at $E = 0$ by adding $eEd/\epsilon_r^*$ to the on-site energies for the top layer and $-eEd/\epsilon_r^*$ to the on-site energies for the bottom layer. The evolution with $E$ of the difference in valence band maxima $E_\Gamma - E_K$ for $\mathbf{k} \cdot \mathbf{p}$ models obtained in this way is shown by the solid red line in Fig. 4(c).

The origin of the different $E$-field dependences at $K$ and $\Gamma$ is the difference in orbital character of the corresponding eigenstates. In particular, the valence band maximum states at $K$ have opposite angular momentum and hence opposite spin in outside and interior layers, which suppresses hybridization between adjacent layers. At $E = 0$, the highest- and lowest-energy states among the top group of three valence states at $K$ are dominated by amplitudes on the top and bottom layers combined with even and odd parity, respectively. These subbands have dominant $d$-orbital amplitudes with $|l_z = -2, \downarrow\rangle$ representations on the top and bottom layers, and a much smaller $|l_z = +2, \downarrow\rangle$ amplitude on the middle layer. Importantly, because of the change in the sign of $l_z$, the on-site energy on the middle layer is shifted to lower energy by the spin-orbit interaction, explaining its small amplitude in the high energy subbands. The middle state among the three high energy subbands is dominated by a middle layer $d$-orbital amplitude with the same $l_z$ character as its outer layer cousins, but opposite spin. Like the outside layer subbands, it is localized in the middle layer by spin-orbit splitting of the $\uparrow$ $d$-orbital on-site energies. As $E$ is increased, the eigenstate character of the states at $K$ is retained, leading to strong localization of subband wavefunctions on individual layers for $E \gtrsim 0.1$ V/nm. At $\Gamma$, on the other hand, the eigenstate amplitudes include similar $l_z = 0$ representations on all layers, allowing for large effective interlayer coupling and explaining a greatly reduced rate of change of the eigenstates and their energies with increasing $E$.

To compare the theoretical results with the experimental data of Fig. 3(c), we consider the equilibrium distribution at a finite hole density. We calculate the Fermi energy for a given total hole density using an effective-mass parameterization of the highest valence bands. Holes are distributed among layers and valleys accounting for screening self-consistently with a simple electrostatic model, considering a planar charge distribution at each W layer. The variation of $E_\Gamma - E_K$ with $E$ is modified in the presence of holes, as shown by the black circles and line in Fig. 4(c). The distribution of holes between the $\Gamma$ and $K$ valleys as a function of $E$ is shown in Fig. 4(d), displaying a trend which is qualitatively consistent with the measured occupations of Fig. 3(c). We obtain a hole effective mass at $\Gamma$ ($K$) of $m_\Gamma^* = 0.82 m_e$ ($m_K^* = 0.35 m_e$). These $m^*$ values are smaller than the measured masses, but they have a similar ratio. Considering the masses difference, the transfer of holes from $\Gamma$ to $K$ in our model is expected to begin at a smaller $E$-field and proceed more slowly with increasing $E$, both features seen by comparing Fig. 3(c) and Fig. 4(d) data. The ∼50% reduction of the calculated hole transfer rate by comparison to experiments is larger than the ∼30% reduction that would be expected due to the smaller calculated effective masses, and the corresponding reduction in density of states. The remaining difference may be attributable to the neglect of interaction effects beyond the Hartree level, and the simplicity of the screening model. In particular, exchange interactions are expected to further increase $E_K$ as these orbitals become occupied, and to enhance the rate of change of $E_\Gamma - E_K$ with $E$-field.

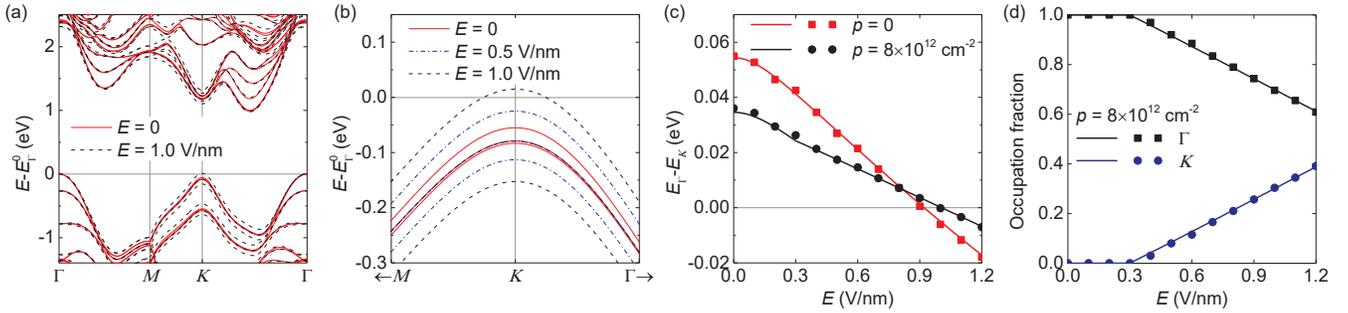

FIG. 4. (a) Band structure of trilayer $2H-WSe_2$ with and without $E$-field. Energies are given relative to the valence band maximum at $\Gamma$ at $E = 0$. (b) Zoomed-in view of the valence band maximum at $K$, showing an approximately linear shift with $E$-field. (c) $E$-field-dependence of $E_\Gamma - E_K$, at zero and a finite hole density. (d) $E$-field dependence of the relative occupation of the $\Gamma$ and $K$ valleys at a fixed $p = 8 \times 10^{12}$ cm$^{-2}$. In panels (c) and (d), the symbols (lines) show results where the $E$-field is included via the DFT calculation ($\mathbf{k} \cdot \mathbf{p}$ model).

Figure 2(b) data can be explained as follows. At low $|V_{TG}|$ a majority of the holes reside in the $\Gamma$ valley, thanks to a large $E_\Gamma - E_K$ at low $E$-fields. Increasing $|V_{TG}|$ increases the $E$-field and the total density, while decreasing $E_\Gamma - E_K$. The concomitant density increase and $E_\Gamma - E_K$ decrease result in a nearly constant $\Gamma$ valley density.

In summary, we observe and explain exceptional sensitivity of $\Gamma - K$ valley hole populations in trilayer WSe$_2$ to applied transverse electric fields. Given the distinct properties of holes in the two valleys characterized by differences in effective mass, spatial localization, and Berry curvature [21], the $E$-field can be used as an effective knob to tune the electronic and spin transport properties in trilayer WSe$_2$, thereby making it an interesting system for valleytronics [1, 21]. As an example, trilayer WSe$_2$ is a unique system to tune the spin Hall effect [22] by transferring carriers from the $\Gamma$ to the $K$ valley using a transverse electric field.

This work was supported by the Nanoelectronics Research Initiative SWAN center, and Intel Corp. T. L. and A. H. M. acknowledge support by the Department of Energy, Office of Basic Energy Sciences under Contract No. DE-FG03-02ER45958 and by the Welch Foundation under Grant No. TBF1473. The authors acknowledge the Texas Advanced Computing Center (TACC) at The University of Texas at Austin for providing HPC resources that have contributed to the research results reported within this paper. A portion of this work was performed at the National High Magnetic Field Laboratory, which is supported by National Science Foundation Cooperative Agreement No. DMR-1157490, and the State of Florida.

H. C. P. M. and T. L. contributed equally to this work.

---

* etutuc@mer.utexas.edu

# Supplemental Material:
# Tunable $\Gamma - K$ Valley Populations in Hole-Doped Trilayer WSe$_2$

Hema C. P. Movva,[1] Timothy Lovorn,[2] Babak Fallahazad,[1] Stefano Larentis,[1] Kyounghwan Kim,[1]
Takashi Taniguchi,[3] Kenji Watanabe,[3] Sanjay K. Banerjee,[1] Allan H. MacDonald,[2] and Emanuel Tutuc[1, *]

[1]*Microelectronics Research Center, Department of Electrical and Computer Engineering,
The University of Texas at Austin, Austin, TX 78758, USA*
[2]*Department of Physics, The University of Texas at Austin, Austin, TX 78712, USA*
[3]*National Institute for Materials Science, 1-1 Namiki Tsukuba, Ibaraki 305-0044, Japan*
(Dated: January 7, 2018)

## SdH OSCILLATIONS $V_{BG}$ DEPENDENCE

Figure S1(a) shows $R_{xx}$ vs $B$ measured in the same trilayer WSe$_2$ sample discussed in the main text at different $V_{BG}$ values, $V_{TG} = -12$ V, and $T = 1.5$ K. Figure S1(b) shows the FT amplitude vs frequency corresponding to the $R_{xx}$ vs $B^{-1}$ data of Fig. S1(a). The FT data consist of one or two characteristic peak(s) depending on the value of $V_{BG}$. When $V_{BG} \leq 0$ V, the FT data show two characteristic peaks analogous to Fig. 2(b) in the main text. However, the low-frequency peak increases in frequency and the high-frequency peak decreases in frequency with increasing $V_{BG}$. The two peaks merge into one broad peak at $V_{BG} = 1$ V and cannot be individually resolved. For $1$ V $\leq V_{BG} \leq 4$ V, the resulting peak increases in frequency with increasing $V_{BG}$. When $V_{BG} \geq 4$ V, this peak reverses its trend and starts decreasing in frequency with increasing $V_{BG}$. This unusual $V_{BG}$ dependence suggests depopulation of one subband with increasing $V_{BG}$.

We measured the SdH oscillations $T$-dependence where the FT data show only one peak to identify the depopulated subband. Figure S1(c) shows $R_{xx}$ vs $B$ at $V_{TG} = -12$ V, $V_{BG} = 6$ V, and various $T$ values. Using the same technique described in the main text, we extracted an $m^* = 0.5m_e$ associated with the sole FT peak [Fig. 2(e) in the main text; $p = 3.0 \times 10^{12}$ cm$^{-2}$]. The same $m^*$ as the low-frequency peak at $V_{TG} = -12$ V, $V_{BG} = 0$ V [Fig. 2(e) in the main text; $p = 8.0 \times 10^{12}$ cm$^{-2}$] implies that both these peaks belong to the same subband, namely, the $K$ valley. The high-frequency peak at $V_{BG} \leq 0$ V therefore belongs to the $\Gamma$ valley. Figure S1(d) summarizes $p$ vs $V_{BG}$ determined from the FT peaks of Fig. S1(b) data ($\Gamma$, $K$) using $g = 2$, their sum ($\Gamma + K$), and from Hall measurements. The close agreement between the total FT and Hall densities corroborates our interpretation. We note that in the range $1$ V $\leq V_{BG} \leq 4$ V, the component peak frequencies cannot be individually resolved, and are therefore omitted.

## DENSITY FUNCTIONAL THEORY CALCULATIONS

We perform fully-relativistic density functional theory (DFT) calculations under the local-density approximation (LDA) [S1] for the WSe$_2$ trilayer system using the QUANTUM ESPRESSO distribution [S2]. We choose the LDA based on previous work which shows that it provides a reasonable de-scription of bilayer graphene and graphite when compared to more sophisticated and computationally-intensive approaches, including quantum Monte Carlo and the random phase approximation [S3, S4], which suggests adequate performance of the LDA for describing the interlayer interaction in van der Waals-bonded materials more broadly. The qualitative agreement of our result for the valley occupations, shown in Fig. 4(d) in the main text, with the observed occupations, shown in Fig. 3(c) in the main text, appears to justify the choice of LDA. To explore the robustness of this result, we performed the calculation again under the Perdew-Burke-Ernzerhof (PBE) exchange-correlation functional [S5]; the result of this calculation is shown in Fig. S2, given in the same way as Fig. 4(c) and (d) in the main text. Under PBE, we find effective masses which are almost identical to those in LDA, but $E_\Gamma - E_K$ at $E = 0$ is significantly reduced (down to $\approx 0.04$ eV at $p = 0$ versus $\approx 0.055$ eV at $p = 0$ for LDA, and $\approx 0.02$ eV versus $\approx 0.035$ eV at $p = 8 \times 10^{12}$ cm$^{-2}$). This reduction in $E_\Gamma - E_K$ leads to a small fraction ($\approx 5\%$) of holes occupying the $K$ valley at $E = 0$ for $p = 8 \times 10^{12}$ cm$^{-2}$. The effective dielectric constant $\epsilon_r^*$ is reduced slightly in the PBE case (7.57 in PBE versus 7.87 in LDA). For $E \gtrsim 0.1$ V/nm, $E_\Gamma - E_K$ decreases linearly with increasing $E$, and the $K$ valley occupation increases linearly with increasing $E$ at approximately the same rate as the LDA case.

We use norm-conserving pseudopotentials based on those provided by the SG15 pseudopotential library [S6]; this library provides optimized inputs for the ONCVPSP pseudopotential generation method [S7]. We alter the pseudopotential generation inputs from the SG15 library to produce fully-relativistic, LDA pseudopotentials; the cutoff radius of the W $p$ states is also reduced slightly to eliminate a ghost state which appears when converting the pseudopotential to fully-relativistic LDA.

We choose structures for the WSe$_2$ layers following the structures reported for monolayers in Ref. [S8], which are obtained by relaxation in DFT with the PBE exchange-correlation functional [S5]. We take the interlayer distance to be equal to that of bulk WSe$_2$ as given by Ref. [S9] and choose a 20 Å vacuum distance for the slab model. We choose a plane-wave cutoff energy of 60 Ry and corresponding charge density cutoff energy of 240 Ry, and we sample the Brillouin zone with an $18 \times 18 \times 1$ grid of $k$-points. The calculations are well-converged with respect to cutoff energy and $k$-point sampling; we obtain very similar results when the plane-wave and charge density cutoff energies are increased to 80 Ry and 320 Ry, respectively, as well as when the number of $k$-points is reduced to $9 \times 9 \times 1$.

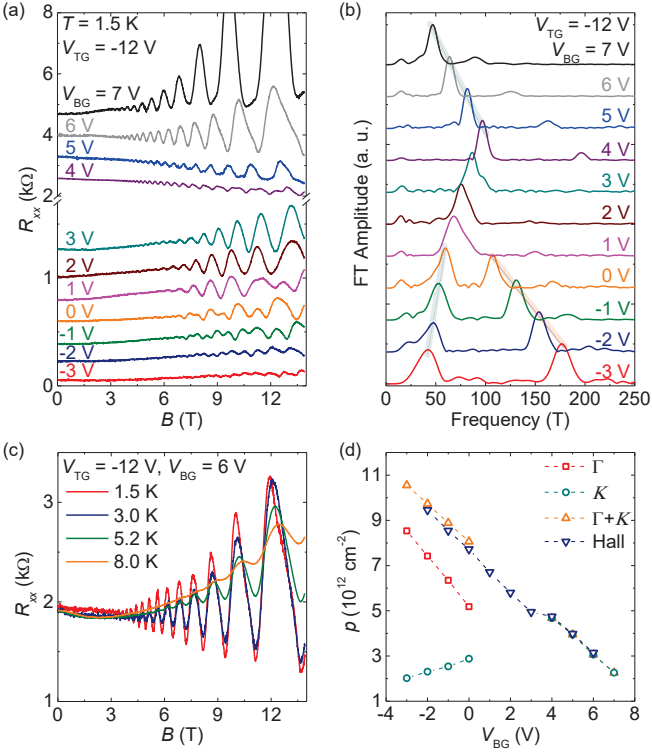

FIG. S1. (a) $R_{xx}$ vs $B$ at various $V_{BG}$ values, $V_{TG} = -12$ V, and $T = 1.5$ K; the traces are shifted for clarity. (b) FT amplitude vs frequency corresponding to the $R_{xx}$ vs $B^{-1}$ data of panel (a); the traces are shifted for clarity. When $V_{BG} \leq 0$ V, the low-frequency peak increases in frequency (dashed line guide to the eye) and the high-frequency peak decreases in frequency with increasing $V_{BG}$ (solid line guide to the eye). For $1$ V $\leq V_{BG} \leq 4$ V, the two peaks merge into one single peak whose frequency increases with increasing $V_{BG}$. When $V_{BG} \geq 4$ V, the sole peak starts decreasing in frequency with increasing $V_{BG}$ (dashed line guide to the eye). (c) $R_{xx}$ vs $B$ at $V_{TG} = -12$ V, $V_{BG} = 6$ V measured at different $T$ values. (d) $p$ vs $V_{BG}$ associated with the individual FT peaks of panel (b) data ($\Gamma$, $K$), their sum ($\Gamma + K$), and Hall measurements. Only data points where the component peak frequencies can be clearly resolved are shown.

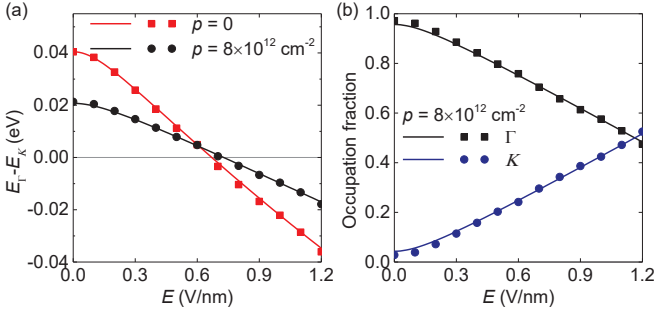

FIG. S2. PBE result for (a) $E_\Gamma - E_K$ versus $E$ and (b) valley occupations versus $E$. Data is plotted in the same way as the LDA result shown in Fig. 4(c) and (d) in the main text.

The transverse electric field $E$ is included in the DFT calculation by the addition of a sawtooth potential (giving an electric field $E$ in the trilayer WSe$_2$ region, and an opposing field in a small part of the vacuum region as required to keep the potential periodic). This potential is modified by the dipole correction of Ref. [S10], as required due to the dipole induced in the trilayer WSe$_2$ by $E$.

We obtain the Hamiltonian in a tight-binding basis using WANNIER90 [S11]. We project onto $p$ orbitals for Se and $d$ orbitals for W, following Ref. [S12], and reproduce the DFT bandstructure with high accuracy. We choose an outer window for disentanglement 10 eV below the Fermi level and 5 eV above it, and an inner window 8 eV below the Fermi level and 3 eV above it. To ensure that the identity of the orbitals is maintained, we do not perform maximal localization (in particular, we want to associate each orbital with a particular WSe$_2$ layer).

The software used to generate inputs for QUANTUM ESPRESSO and WANNIER90, and to obtain the $\mathbf{k} \cdot \mathbf{p}$ model and calculate layer occupations as discussed in the following sections, is available at https://github.com/tflovorn/displ. A dataset including these inputs and associated outputs, as well as the full $\mathbf{k} \cdot \mathbf{p}$ model, is available at https://doi.org/10.5281/zenodo.1135279.

## $\mathbf{k} \cdot \mathbf{p}$ MODEL

Using the tight-binding models obtained by WANNIER90 at each transverse electric field value, we derive corresponding $\mathbf{k} \cdot \mathbf{p}$ models. The $\mathbf{k} \cdot \mathbf{p}$ models we consider are restricted to the bands and $k$-space regions of interest - the top 6 valence bands near the $\Gamma$ and $K$ points - and represented in a basis of layer-projected orbitals. As the first step in this process, we construct layer projection operators:

$$P_l = \sum_{i \in W_l} |i\rangle \langle i| . \quad (S1)$$

Here $W_l$ is the set of orbitals in the Wannier tight-binding basis which are located on the WSe$_2$ layer $l$. Then, considering the Hamiltonian eigenstates $|k_0 m\rangle$, where $k_0 = \Gamma$, $K$, or $K'$ and the band index $m$ is restricted to the top 6 valence states, we construct the layer-projected density matrix for these states:

$$\rho_l = \sum_m P_l |k_0 m\rangle \langle k_0 m| P_l. \quad (S2)$$

The optimal choices for the layer-projected eigenstates are then obtained by choosing the 2 eigenstates (6 states / 3 layers) of the density matrix with the largest eigenvalues. In the following, we will refer to these states as $|li\rangle$. If the layer-resolved representation perfectly captures the Hamiltonian eigenstates $|k_0 m\rangle$, the layer-projected density matrix will have 2 eigenvalues with value 1 and the remaining eigenvalues will have value

0. For $E = 0$, we find a representation at $K$ where the largest two eigenvalues on each layer are all greater than 0.99; at $\Gamma$, the accuracy of the representation is weaker, with the largest two eigenvalues on each layer being $\approx 0.92$. Similar quality of representation is found at $E = 1.2$ V/nm.

The $\mathbf{k} \cdot \mathbf{p}$ Hamiltonian is expressed as:

$$H_{k_0}(k + k_0) = H_0 + H_0' + \sum_a (p_a + p_a')k_a$$
$$- \sum_{a'a} \left( \frac{\hbar^2}{2m_{a'a}^*} + \frac{\hbar^2}{2m_{a'a}^{*'}} \right) k_{a'} k_a \quad (S3)$$

Here the unprimed terms are obtained directly from the basis derived above, and the primed terms are corrections obtained by Löwdin quasi-degenerate perturbation theory [S13] which include the effect of states present in the full tight-binding basis but not included in this basis; each of these terms has a $k_0$ dependence which has been suppressed. The sums over $a, a'$ run over Cartesian components. $H_0$ is obtained from the expression

$$[H_0]_{l'i',li} = \langle l'i'|H_{k_0}^W|li \rangle \quad (S4)$$

where $H_k^W$ is the tight-binding Hamiltonian in the Wannier basis. Similarly, the momentum and effective mass terms are obtained from

$$[p_a]_{l'i',li} = \langle l'i'| \left. \frac{\partial H_k^W}{\partial k_a} \right|_{k_0} |li \rangle \quad (S5)$$

and

$$\left[ \frac{-\hbar^2}{m_{a'a}^*} \right]_{l'i',li} = \langle l'i'| \left. \frac{\partial^2 H_k^W}{\partial k_{a'} \partial k_a} \right|_{k_0} |li \rangle. \quad (S6)$$

To obtain the correction, we decompose the tight-binding basis into states in the $|li\rangle$ basis, which we denote as $P$, and states in the complement of the basis, which we denote as $Q$. The tight-binding Hamiltonian is then expressed as blocks $H_{PP}$, $H_{PQ}$, $H_{QP} = H_{PQ}^\dagger$, and $H_{QQ}$. $H_{PP}$ is the same as $H_0$ above. $H_{PQ}$ and $H_{QQ}$ are given by

$$\left[ H_{PQ}(k) \right]_{li,j} = \langle li|H_k^W|j \rangle, \quad (S7)$$

and

$$\left[ H_{QQ}(k) \right]_{j',j} = \langle j'|H_k^W|j \rangle \quad (S8)$$

where $j', j \in Q$. The perturbative correction takes the form

$$H'(k) = H_{PQ}(k) \left[ E_{rep}\mathbb{1}_{QQ} - H_{QQ}(k) \right]^{-1} H_{QP}(k). \quad (S9)$$

Here $E_{rep}$ is a representative energy which is near the energy of all states in P. We choose $E_{rep}$ equal to the highest valence band energy at $k_0$. The correction terms appearing in the $\mathbf{k} \cdot \mathbf{p}$ Hamiltonian are then obtained using $H'(k)$ as

$$[H']_{l'i',li} = \langle l'i'|H'(k_0)|li \rangle, \quad (S10)$$

$$[p_a']_{l'i',li} = \langle l'i'| \left. \frac{\partial H'(k)}{\partial k_a} \right|_{k_0} |li \rangle, \quad (S11)$$

and

$$\left[ \frac{-\hbar^2}{m_{a'a}^{*'}} \right]_{l'i',li} = \langle l'i'| \left. \frac{\partial^2 H'(k)}{\partial k_{a'} \partial k_a} \right|_{k_0} |li \rangle, \quad (S12)$$

To illustrate the structure of the layer-resolved $\mathbf{k} \cdot \mathbf{p}$ model and the related structure of the eigenstates, here we explicitly describe the $H_0$ terms at $K$ and $\Gamma$ for $E = 0$. In both the $K$ and $\Gamma$ cases, the basis states include only the bottom layer for the first two states, only the middle layer for the middle two states, and only the top layer for the last two states. The basis states at $K$ are dominated by contributions from the W $d$-orbital $l_z = \pm 2$ states and the Se $p$-orbital $l_z = \pm 1$ states; combining these contributions as ± and indicating the spin, the basis has the form: $(-,\uparrow)$, $(-,\downarrow)$, $(+,\downarrow)$, $(+,\uparrow)$, $(-,\uparrow)$, $(-,\downarrow)$. The maximum weight outside this orbital decomposition is 1.2%. The $H_0$ terms at $K$ (in units of eV, with values below 1 meV truncated to 0, and with the zero of energy shifted to the valence band maximum at $\Gamma$) are:

$$H_{0,K} + H_{0,K}' = \begin{pmatrix} -0.544 & 0 & 0 & 0.064 & 0.006 & 0 \\ 0 & -0.077 & 0.064 & 0 & 0 & 0.006 \\ 0 & 0.064 & -0.562 & 0 & 0 & 0.064 \\ 0.064 & 0 & 0 & -0.097 & 0.064 & 0 \\ 0.006 & 0 & 0 & 0.064 & -0.544 & 0 \\ 0 & 0.006 & 0.064 & 0 & 0 & -0.077 \end{pmatrix} \quad (S13)$$

The difference in on-site energies of the basis states at $K$ is dominated by the spin-orbit coupling $L_z S_z$ term. Due to their difference in spins, the second and sixth basis states are decoupled from the fourth basis state (decoupled within the high-energy group); the first and fifth basis states are similarly decoupled from the third basis state (low-energy group). The weak coupling between the high- and low-energy groups becomes a much weaker effective coupling within the groups due to the difference in energies generated by the spin-orbit splitting. The direct coupling between the second and sixth and first and fifth basis states, i.e. from the bottom to the top layer, is also weak.

The basis states at $\Gamma$ are dominated by contributions from $l_z = 0$ states on both the W $d$-orbitals and Se $p$-orbitals, with the maximum weight outside this orbital decomposition being 5.5%. Due to the Kramers degeneracy, the basis states are not spin-polarized in a generic sense (although the basis states on each layer, being eigenstates of $\rho_l$ with the same eigenvalues, may be mixed via a unitary transformation to enhance their spin polarization; no such special representation has been chosen here). The $H_0$ terms at $\Gamma$ (formatted in the same way as those at $K$) are:

$$H_{0,\Gamma} + H'_{0,\Gamma} = \begin{pmatrix} -0.395 & 0 & -0.194+0.017i & 0.193+0.067i & 0.017+0.001i & -0.099+0.001i \\ 0 & -0.395 & 0.203+0.027i & 0.165+0.102i & -0.099+0.001i & -0.017+0.001i \\ -0.194-0.017i & 0.203-0.027i & -0.321 & 0 & 0.232-0.024i & -0.157-0.023i \\ 0.193-0.067i & 0.165-0.102i & 0 & -0.321 & 0.129-0.093i & 0.217-0.086i \\ 0.017-0.001i & -0.099-0.001i & 0.232+0.024i & 0.129+0.093i & -0.395 & 0 \\ -0.099-0.001i & -0.017-0.001i & -0.157+0.023i & 0.217+0.086i & 0 & -0.395 \end{pmatrix}$$
$$\text{(S14)}$$

In this case, adjacent layers are strongly coupled, and a substantial direct coupling between bottom and top layers is also present. The $L_z S_z$ splitting and spin polarization which suppress effective interlayer hopping are not present at $\Gamma$. The strong interlayer coupling may also be linked directly to the orientation of the $l_z = 0$ orbitals, which could have substantial interlayer overlap (we have not investigated this directly).

A schematic picture of the structure of the basis states of the layer-resolved $\mathbf{k} \cdot \mathbf{p}$ model at $K$ and $\Gamma$ and the resulting structure of the eigenstates is shown in Fig. S3. The basis states shown in Fig. S3(a) reflect the energies (diagonal elements) and dominant hopping terms given in the $H_0$ matrices above, with basis states at $K$ split by spin-orbit coupling into high- and low-energy groups. The resulting eigenstates, shown in Fig. S3(b), display distinct energy-splitting mechanisms at $K$ and $\Gamma$. The eigenstates at $K$ remain split primarily by spin-orbit coupling, with states within the high- and low-energy groups additionally split by weak coupling between bottom and top layers (via effective coupling through the middle layer as well as direct bottom-to-top coupling). The eigenstates at $\Gamma$ are split by the strong coupling between adjacent layers, unhindered by the small (compared to the interlayer coupling) energy difference between layer basis states. Here the factor of $2^{3/2}$ associated with the splitting at $\Gamma$ is obtained from the eigenvalues $\pm\sqrt{2}t$ of a $3 \times 3$ matrix with zeros on the main diagonal and equal hopping values $t$ on the diagonals above and below the main diagonal.

# ELECTRIC FIELD RESPONSE

We calculate an effective $c$-axis dielectric constant $\epsilon_r^*$ using layer-resolved $\mathbf{k} \cdot \mathbf{p}$ models derived at various values of the transverse electric field $E$ (which has been included through the DFT calculation). The effective dielectric constant for

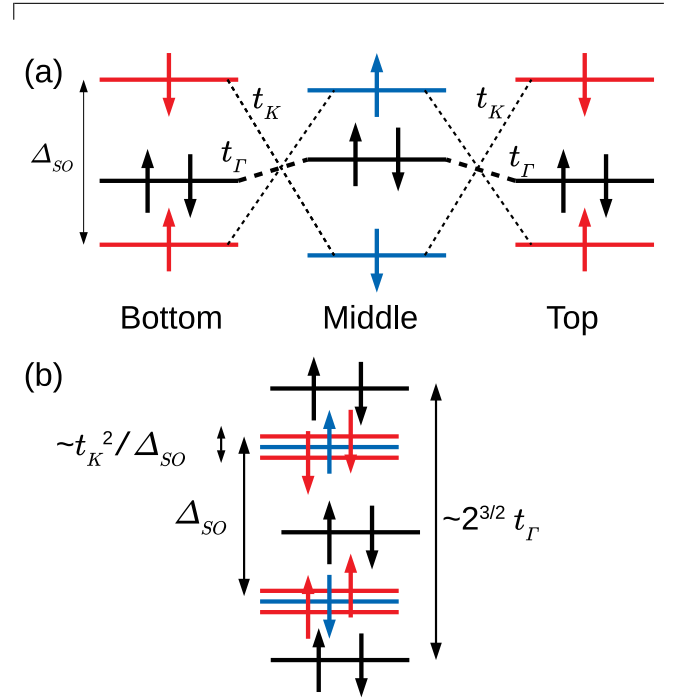

FIG. S3. Schematic structure of the states at $K$ and $\Gamma$. (a) Basis states on each layer of the layer-resolved $\mathbf{k} \cdot \mathbf{p}$ model. States at $K$ are at high and low energies, split by spin-orbit coupling, with $l_z < 0$-dominated states colored red and $l_z > 0$-dominated states colored blue. States at $\Gamma$, with $l_z = 0$, are colored black. (b) Eigenstates at $K$ and $\Gamma$. Energy splittings are described in the text.

states in valley $v$ ($v = \Gamma$ or $K$) is given by

$$E_{v,\text{top}} - E_{v,\text{bottom}} = 2eEd/\epsilon_{r,v}^*. \tag{S15}$$

Here $E_{v,\text{top}}$ is the "on-site" energy for the top layer in valley $v$, corresponding to a term on the diagonal in the k-independent part of the $\mathbf{k} \cdot \mathbf{p}$ Hamiltonian in which states have been projected to the top layer; $E_{v,\text{bottom}}$ is defined in the same way for the bottom layer. For the $K$ valley, where there are two inequivalent choices of on-site energy for each layer (corresponding to $L_z S_z = \pm 1$ states), the value of $\epsilon^*_{r,K}$ obtained is very similar for either choice; we choose the higher-energy group of basis states for calculation of $\epsilon^*_{r,K}$. Averaging the values obtained in this way over electric field choices in the range $[0.5, 1.2]$ V/nm at 0.1 V/nm intervals, we obtain $\epsilon^*_{r,K} = 7.69$ and $\epsilon^*_{r,\Gamma} = 8.05$ (electric field values below 0.5 V/nm are excluded due to exhibiting larger variation of effective dielectric constant with electric field). As discussed in the main text, we choose the average of these two values, and we construct $\mathbf{k} \cdot \mathbf{p}$ models at finite $E$ from the model at $E = 0$ by adding corresponding terms to the on-site energies.

When the hole density is finite, there is an additional contribution to the layer energies due to the electrostatic potential of holes on each layer. The hole distribution will rearrange to screen the electric field: we must calculate the distribution of holes among the layers and the layer Hamiltonian self-consistently. We calculate the density of holes on each layer $\frac{1}{A} n_h^l(E_F)$ using a further simplification of the $\mathbf{k} \cdot \mathbf{p}$ model. Specifically, given a value of the electrostatic potential $\phi$ due to the applied transverse electric field and the holes, we construct the corresponding $\mathbf{k} \cdot \mathbf{p}$ Hamiltonian $H^\phi_{\mathbf{k} \cdot \mathbf{p}}(k)$. Then we represent this Hamiltonian as a set of decoupled bands with energies

$$\epsilon_{nk;k_0} \approx E_n(k_0) - \frac{\hbar^2}{2m^*_{n;k_0}}(k_x^2 + k_y^2). \quad (S16)$$

Here $E_n(k_0)$ are the eigenvalues of $H^\phi_{\mathbf{k} \cdot \mathbf{p}}(k_0)$ and the effective masses are obtained by numerical differentiation of the eigenvalues around $k_0$. In this representation, the density of holes on each layer is given by

$$\frac{1}{A} n_h^l(E_F) = \frac{1}{4\pi} \sum_{k_0 n} |P_l |n_{k_0}\rangle|^2 \left(\frac{\hbar^2}{2m^*_{n;k_0}}\right)^{-1}$$
$$\cdot (E_n(k_0) - E_F)\Theta(E_n(k_0) - E_F). \quad (S17)$$

Here $P_l |n_{k_0}\rangle$ is the projection of the $n$'th eigenstate at $k_0$ (as defined by its energy $\epsilon_{nk;k_0}$) onto the layer $l$. The occupation of each valley is given by evaluating each $k_0$ term here separately.

The total density of holes is given by summing over layers, and the Fermi energy $E_F$ is obtained by solving for the $E_F$ which gives the correct total density of holes. Electrostatic doping of transition metal dichalcogenides has previously been studied theoretically up to much higher doping levels than we consider here using a more sophisticated method, but without the application of an independent transverse electric field (i.e. in the context of a single-gated rather than dual-gated device) [S14].

The electrostatic potential due to holes is obtained by taking the hole density to be distributed in uniform sheets centered on the W atoms. With this charge distribution, the contribution to the electrostatic potential due to the holes is $\frac{-d}{2\epsilon^*_r \epsilon_0}(\sigma_2 + 2\sigma_3)$ on the bottom layer, $\frac{-d}{2\epsilon^*_r \epsilon_0}(\sigma_1 + \sigma_3)$ on the middle layer, and $\frac{-d}{2\epsilon^*_r \epsilon_0}(2\sigma_1 + \sigma_2)$ on the top layer, where $\sigma_l = \frac{e}{A} n_h^l(E_F)$ is the charge density on layer $l$.

---

* etutuc@mer.utexas.edu